**Quantifying Circadian Desynchrony in ICU Patients and Its Association with Delirium**

Yuanfang Ren[1,2], Andrea E. Davidson[1,2], Jiaqing Zhang[3], Miguel Contreras[4], Ayush K. Patel[1,2], Michelle Gumz[2], Tezcan Ozrazgat-Baslanti[1,2], Parisa Rashidi[1,4]*, Azra Bihorac[1,2]*

* These senior authors have contributed equally

[1] Intelligent Clinical Care Center (IC3), University of Florida, Gainesville, FL, USA

[2] Division of Nephrology, Hypertension, and Renal Transplantation, Department of Medicine, College of Medicine, University of Florida, Gainesville, FL, USA

[3] Department of Electrical and Computer Engineering, University of Florida, Gainesville, FL, USA

[4] J. Crayton Pruitt Family Department of Biomedical Engineering, University of Florida, Gainesville, FL, USA

**Corresponding author:** Azra Bihorac MD, MS, Department of Medicine, Division of Nephrology, Hypertension, and Renal Transplantation, PO Box 100224, Gainesville, FL 32610-0224. Telephone: (352) 294-8580; Fax: (352) 392-5465; Email: abihorac@ufl.edu

**Abstract**

**Background:** Circadian desynchrony characterized by the misalignment between an individual's internal biological rhythms and external environmental cues, significantly affects various physiological processes and health outcomes. Quantifying circadian desynchrony often requires prolonged and frequent monitoring, and currently, an easy tool for this purpose is missing. Additionally, its association with the incidence of delirium has not been clearly explored.

**Methods:** A prospective observational study was carried out in intensive care units (ICU) of a tertiary hospital. Circadian transcriptomics of blood monocytes from 86 individuals were collected on two consecutive days, although a second sample could not be obtained from all participants. Using two public datasets comprised of healthy volunteers, we replicated a model for determining internal circadian time. We developed an approach to quantify circadian desynchrony by comparing internal circadian time and external blood collection time. We applied the model and quantified circadian desynchrony index among ICU patients, and investigated its association with the incidence of delirium.

**Results:** The replicated model for determining internal circadian time achieved comparable high accuracy. The quantified circadian desynchrony index was significantly higher among critically ill ICU patients compared to healthy subjects, with values of 10.03 hours vs 2.50-2.95 hours ($p < 0.001$). Most ICU patients had a circadian desynchrony index greater than 9 hours. Additionally, the index was lower in patients whose blood samples were drawn after 3pm, with values of 5.00 hours compared to 10.01-10.90 hours in other groups ($p < 0.001$). The distribution of clinical characteristics and outcomes across groups categorized by circadian desynchrony showed no statistical significance. Although no statistically significant differences in delirium rates were found across different circadian desynchrony levels, trends suggest that higher desynchrony indices may be associated with a greater propensity for delirium.

**Conclusions:** we developed a simple approach to quantify circadian desynchrony index among ICU patients, requiring only a single blood sample taken any time during the day. While no statistically significant differences in delirium rates were observed across different levels of circadian desynchrony, the trends indicate that patients with higher desynchrony indices may be more prone to delirium. This highlights the need for further research to explore these associations in greater depth.

**Introduction**

Circadian desynchrony characterized by the misalignment between an individual's internal biological rhythms and external environmental cues, significantly affects various physiological processes and health outcomes.[1] In recent years, it has gained increased attention due to its implications in numerous health issues, such as sleep disorders, metabolic dysfunction, cognitive impairment, and increased susceptibility to diseases.[2] Given the clinical relevance, accurately quantifying circadian desynchrony is crucial for understanding its impact, particularly in sensitive populations such as critically ill patients.

Several methods have been developed and widely adopted for quantifying circadian desynchrony.[3] Actigraphy, the most commonly applied method in human studies, employs wrist-worn devices that record movement patterns, enabling assessment of sleep-wake cycles through the analysis of activity onset and offset timing.[4] Sleep diaries provide subjective records of sleep and wake times, allowing calculation of sleep midpoints to infer circadian misalignment.[5] Core body temperature measurement, involving continuous monitoring to capture typical daily fluctuations, also serves as a circadian phase marker, given its predictable daily rhythm. Meanwhile, the measurement of melatonin levels, typically through salivary sampling at multiple time points, remains widely considered the gold standard for assessing circadian phase.[6] Melatonin secretion is tightly regulated by the internal circadian clock, with its onset serving as a precise biomarker for internal biological timing. Additionally, experimental protocols such as the constant routine, which involves controlling environmental conditions (light, temperature, food intake) to isolate and measure intrinsic circadian rhythms, and the forced desynchrony protocol, which intentionally disrupts circadian rhythms to study their effects, further complement these methods.[7]

Metrics commonly employed to quantify circadian desynchrony include the phase angle, reflecting the timing difference between a measured biological rhythm (e.g., melatonin onset) and a known external reference point (e.g., environmental light exposure), the inter-circadian phase angle, which assesses alignment between two internal rhythms (e.g., activity versus core body temperature), amplitude, indicating rhythm robustness through the difference between peak and trough levels, and circadian period, representing the duration of one complete circadian cycle[4].

Despite their widespread use, these conventional methodologies have notable limitations. Primarily, all these approaches require prolonged and frequent monitoring, making them resource-intensive, intrusive, and challenging to implement in clinical settings such as intensive care units (ICUs).[8] For example, the standard melatonin measurement approach necessitates multiple salivary collections throughout the day, which may be burdensome or impractical for critically ill patients.

Recently, Wittenbrink et al.[9] introduced an innovative method that utilizes a single blood sample to accurately estimate internal circadian time, significantly simplifying circadian assessment. Inspired by this novel approach, we propose using this streamlined method to calculate a circadian desynchrony index, subsequently analyzing its relationship with delirium incidence in critically ill individuals. This approach could potentially overcome previous methodological barriers, providing an efficient, practical, and less intrusive means to study circadian desynchrony in vulnerable patient populations.

**Methods**

*Participants*

Participant recruitment for this prospective observational study took place between January 2021 and July 2023 at a 1000+ bed tertiary hospital in the Southeastern United States.

Eligible participants were adult intensive care unit (ICU) patients who were expected to remain in the ICU for at least 24 hours following consent. Exclusion criteria included patients unable to provide self-consent or have consent provided by a legally authorized representative (LAR), those receiving comfort-only care, and patients placed on contact or isolation precautions. This study received Institutional Review Board (IRB) approval and written informed consent was obtained.

*Sample Collection and Processing*

Blood samples (6-15 mL) were collected from study participants into EDTA tubes by the bedside nurse and immediately handed over to the study team for processing. Collection was attempted on two consecutive days, although a second sample could not be obtained from all participants. Monocytes were isolated from 1-12 mL of whole blood using the EasySep™ Direct Human Monocyte Isolation Kit (Stemcell Technologies), following the manufacturer's protocol. After isolation, the monocytes were lysed with β-mercaptoethanol and stored at -80°C for subsequent RNA extraction. RNA was isolated in batches using the Qiagen RNeasy Micro Kit, following the manufacturer's protocol. RNA quantity and quality were assessed using Qubit Fluorometric Quantification (Invitrogen), Agilent 2100 Bioanalyzer, and/or Agilent 2200 TapeStation. Samples with a total RNA concentration greater than 50 ng/µL were selected for further analysis. The RNA samples that passed quality control were analyzed in batches for gene expression using the NanoString nCounter Digital Analyzer (NanoString Technologies), which generated raw gene transcript counts for a custom set of 13 circadian-related genes. Raw data were normalized for technical variation using the NanoString nSolver software in three steps: (1) background subtraction of the negative controls mean, (2) normalization of the positive controls by the arithmetic mean, and (3) normalization by the geometric mean of the six housekeeping genes. Samples that produced QC flags, mRNA positive normalization flags, and/or mRNA content normalization flags in nSolver were excluded from the final analysis.

*Determination of Internal Circadian Time*

To determine internal circadian time in humans, we replicated the model proposed in the Wittenbrink et al. study[9] using two datasets comprised of healthy volunteers from the same research: the body time (BOTI) study and the validation (VALI) study. The BOTI dataset was employed for model development and internal validation, while the VALI dataset was used for external validation. We subsequently applied the validated model to our own collected data to accurately determine internal circadian time.

To ensure compatibility, we selected 13 common time-telling genes for the model development ('PER2', 'CRY2', 'NR1D1', 'PER1', 'FKBP4', 'HSPH1', 'KLF9', 'CRISPLD2', 'CRY1', 'LGALS3', 'PER3', 'ELMO2', 'NR1D2'). Following the Wittenbrink et al. study, we developed the model using the ZeitZeiger approach and determined the ZeitZeiger's two main parameters sumabsv = {1, 2, 3} and nSPC = {1, 2, 3} using leave-one-subject-out cross validation method.

*Quantifying Circadian Desynchrony Index*

We quantified the circadian desynchrony index using the phase angle difference (PAD) approach. This method measures the temporal difference between internal circadian markers and external time, normalized over a 24-hour cycle, accounting for differences greater than 12 hours using a wrap-around technique. This metric is essential for assessing the synchronization or desynchronization of an individual's internal clock with the external environment. For patients with two samples, we computed the average circadian desynchrony index.

*Outcomes*

The primary outcome was the circadian desynchrony index. Secondary outcomes included the length of ICU and hospital stay, the need for mechanical ventilation, and the incidence of delirium. Delirium was assessed using the Confusion Assessment Method (CAM) score, with a positive CAM score indicating the presence of delirium. The length of stay was calculated for both the entire hospital stay and the period until blood collection. Delirium status was assessed at three points: before blood collection, during the blood collection period (48 hours before the first blood collection and 48 hours after the last blood collection), and throughout the entire hospital stay.

*Statistical Analysis*

We evaluated the performance of our replicated model for determining internal circadian time by calculating the median PAD between the model's predictions and the gold standard dim-light melatonin onset (DLMO) measurement along with the interquartile range (IQR). We assessed the distribution of quantified circadian desynchrony index by calculating the median PAD and IQR, as well as presenting these results in a histogram plot. Additionally, we categorized patients into four groups based on percentiles of the circadian desynchrony index: low, intermediate low, intermediate high, and high. We then compared patient clinical characteristics and clinical outcomes across these patient groups.

For comparison, we performed χ2 test for categorical variables and the Kruskal-Wallis test for continuous variables. The threshold for statistical significance was set a p-value of less than 0.05 for two-sided tests. We adjusted p values for the family-wise error rate due to multiple comparisons using the Bonferroni correction. Statistical analyses were conducted using Python version 3.9 and R version 4.3.1.

## Results

*Performance of Model for Determining Internal Circadian Time*

Table 1 presents the performance of the model across various parameter combinations. Our replicated model achieved its highest accuracy of 0.79 (1.12) hours with the parameters sumabsv=3 and nSpc=3, which is comparable to the best performance reported in the study at 0.8 [1.1] hours. The best model was externally validated on VALI dataset, achieving similar high accuracy as reported in the study (Table 2). Specifically, it achieved 0.76 [0.87] hours vs 0.54 [0.87] hours for morning-collected samples, and 0.49 [0.75] hours vs 0.80 [0.62] hours for afternoon-collected samples.

**Table 1. Performance of internal cross-validation predictors on the development cohort.**

| Parameters of approach | Prediction error, hours, median (IQR) |
|---|---|
| sumabsv = 1, nSpc = 1 | 2.15 (0.78, 7.32) |
| sumabsv = 2, nSpc = 1 | 1.94 (0.77, 7.19) |
| sumabsv = 3, nSpc = 1 | 2.00 (0.87, 7.45) |
| sumabsv = 1, nSpc = 2 | 0.97 (0.48, 1.65) |
| sumabsv = 2, nSpc = 2 | 0.91 (0.31, 1.57) |
| sumabsv = 3, nSpc = 2 | 0.84 (0.36, 1.51) |
| sumabsv = 1, nSpc = 3 | 0.89 (0.47, 1.56) |
| sumabsv = 2, nSpc = 3 | 0.84 (0.28, 1.53) |
| **sumabsv = 3, nSpc = 3** | **0.79 (0.36, 1.48)** |

**Table 2. Performance of predictors on the external validation dataset.**

| Time | Prediction error, hours, median (IQR) | Best accuracy reported in the study, hours, median (IQR) |
|---|---|---|
| morning data | 0.76 (0.37, 1.24) | 0.54 [0.87] |
| afternoon data | 0.49 (0.32, 1.07) | 0.80 [0.62] |

*Distribution of Circadian Desynchrony Index*

Table 3 illustrates the distribution of the circadian desynchrony index, demonstrating that the median circadian desynchrony index is significantly higher in our ICU patients compared to healthy subjects, being 10.03 hours vs 2.50-2.95 hours ($p < 0.001$). Figure 1 further depicts the distribution among ICU patients, demonstrating that the majority had a circadian desynchrony index greater than 9 hours.

**Table 3. Comparison of the circadian desynchrony index among healthy subjects and intensive care unit patients**

| Dataset | circadian desynchrony index, median (IQR), hours |
|---|---|
| BOTI | 2.50 (1.95, 3.52) |
| VALI | 2.95 (1.13, 5.39) |
| ICU patients | 10.03 (8.33, 11.04) |

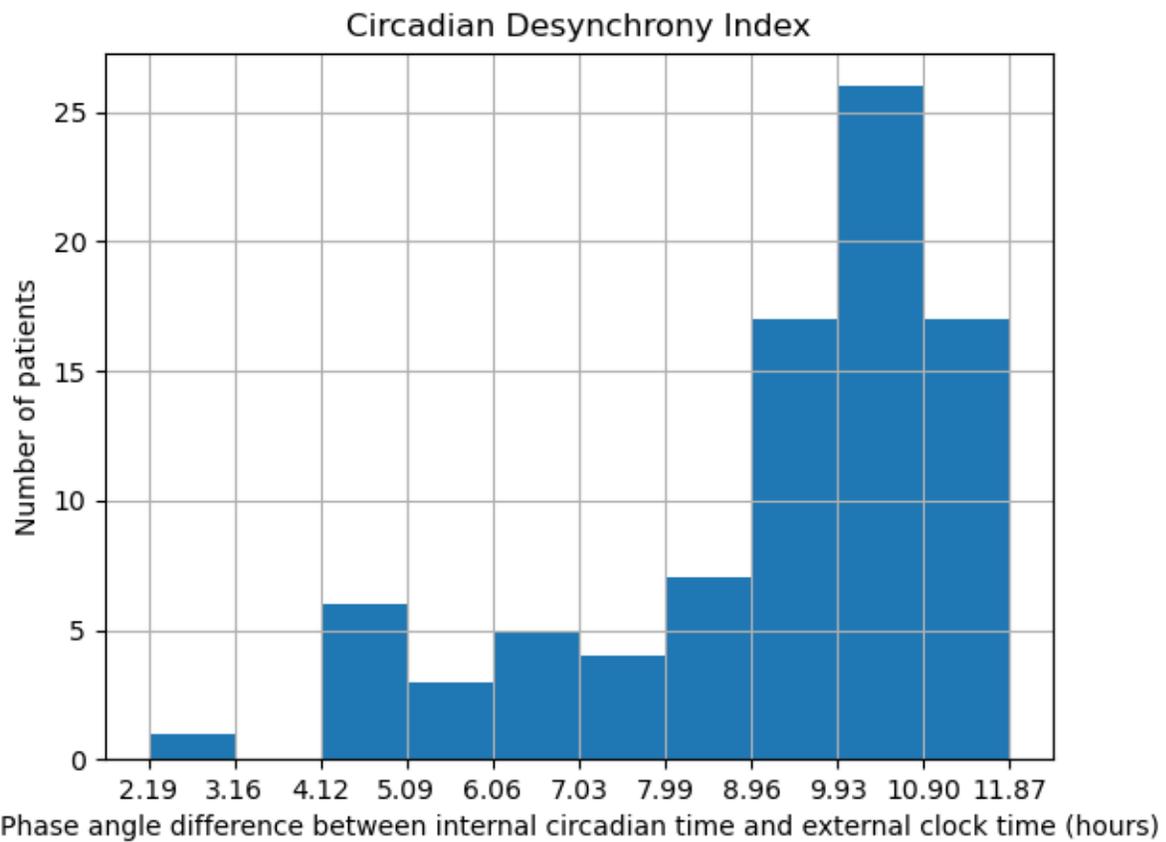

**Figure 1. Distribution of the circadian desynchrony index among intensive care unit patients**

To investigate the impact of sample collection time on the circadian desynchrony index, we grouped our ICU patients based on blood collection time into three categories: <10am,

10am-3pm, and >=3pm. Table 4 presents the distribution of the circadian desynchrony index across these groups. The circadian desynchrony index was lower in the group where blood samples were drawn after 3pm compared to the other groups (5.00 hours vs 10.01-10.90 hours, p < 0.001).

**Table 4. Distribution of circadian desynchrony index among intensive care unit patients stratified by blood sample collection time**

| Blood sample collection time | Circadian desynchrony index, median (IQR), hours |
|---|---|
| <10am (n=24) | 10.90 (10.26, 11.70) |
| 10am-3pm (n=98) | 10.01 (8.83, 10.96) |
| >=3pm (n=13) | 5.00 (4.91, 5.74) |

*Clinical characteristics and outcomes across patients with varying degrees of circadian desynchrony index*

We grouped our ICU patients into four quartiles based on their average circadian desynchrony index: low (2.189 - 8.632 hours), intermediate low (8.632 - 9.915 hours), intermediate high (9.915 - 10.558 hours), and high (10.558 - 11.865 hours). Table 5 presents the clinical characteristics and outcomes for these patient groups.

There were no statistically significant differences in their demographic characteristics across patient groups. The average age of all patients was 58 years (SD 17). Patients in the low and intermediate low circadian desynchrony index groups had a slightly higher average age of 59 years, compared to 57 years in the other groups. Female representation averaged 33%, ranging from 24% to 36% across groups. Most of the patients were White (83%), with African Americans representing 13%, and other races comprising 5% of the cohort. A significant majority (78%) of patients had undergone surgery prior to their last blood collection, and this percentage was consistent across all quartiles. The median number of days between surgery and blood collection was similar across groups, ranging from 2 to 5 days.

There were also no statistically significant differences in their clinical outcomes across patient groups. The median length of ICU stay varied from 10 to 15 days, and the median length of hospital stay ranged from 20 to 23 days among all groups. The median length of ICU stay until blood collection was similar across groups, ranging from 4 to 6 days, and the median length of hospital stay until blood collection ranged from 5 to 7 days. However, patients with the highest circadian desynchrony index had a longer median length of ICU and hospital stay until blood collection compared to other groups. Approximately 42% of patients required mechanical ventilation, with rates ranging from 32% to 52%. Patients with higher circadian desynchrony indices had a greater need for this intervention; with 45% to 52% requiring mechanical ventilation, compared to 32% to 38% in groups with lower desynchrony indices. The incidence of delirium, as indicated by a positive CAM score before the first blood collection, was reported in 12% of patients overall. The breakdown showed 5% in the low circadian desynchrony group, 19% in the intermediate-low group, 14% in the intermediate-high group, and 9% in the high circadian desynchrony group. During the entire hospital stay, 22% of patients had a positive CAM score, with group-specific incidences of 18% (low), 33% (intermediate-low), 14% (intermediate-high), and 23% (high). The most critical period for determining the impact of circadian desynchrony on delirium was within 96 hours of blood collection. During this period, 8% of patients demonstrated a positive CAM score. Notably, no cases of delirium were observed in the low desynchrony group during this time frame. However, 19% of patients in the intermediate-low desynchrony group, 10% in the intermediate-high group, and 5% in the high desynchrony group had positive CAM scores.

**Table 5. Clinical characteristics and outcomes for patients stratified by degree of circadian desynchrony index**

| Variable | All (n=86) | Low (n=22) | Intermediate Low (n=21) | Intermediate High (n=21) | High (n=22) | p-value |
|---|---|---|---|---|---|---|
| Age, years, mean (SD) | 58 (17) | 59 (17) | 59 (16) | 55 (20) | 57 (17) | 0.84 |
| Female sex, n (%) | 28 (33) | 8 (36) | 7 (33) | 5 (24) | 8 (36) | 0.79 |
| Race, n (%) | | | | | | |
| White | 71 (83) | 18 (82) | 20 (95) | 16 (76) | 17 (77) | 0.34 |
| African American | 11 (13) | 3 (14) | 0 (0) | 5 (24) | 3 (14) | 0.15 |
| Other | 4 (5) | 1 (5) | 1 (5) | 0 (0) | 2 (9) | 0.57 |
| Had surgery before the last blood collection | 67 (78) | 17 (77) | 14 (67) | 17 (81) | 19 (86) | 0.46 |
| Days between surgery and blood collection, median (IQR) | 4 (2, 7) | 5 (3, 8) | 4 (2, 6) | 2 (1, 5) | 4 (1, 7) | 0.52 |
| **Outcomes** | | | | | | |
| Length of ICU stay, days, median (IQR) | 12 (6, 27) | 15 (5, 29) | 13 (6, 39) | 10 (6, 16) | 15 (7, 24) | 0.72 |
| Length of hospital stay, days, median (IQR) | 22 (11, 41) | 21 (10, 39) | 21 (8, 42) | 20 (11, 41) | 23 (13, 37) | 0.98 |
| Length of ICU stay until blood collection, days, median (IQR) | 4 (2, 7) | 4 (2, 7) | 4 (1, 5) | 4 (3, 6) | 6 (3, 9) | 0.23 |
| Length of hospital stay until blood collection, days, median (IQR) | 6 (4, 10) | 6 (3, 13) | 5 (2, 7) | 5 (4, 9) | 8 (5, 13) | 0.25 |
| Require mechanical ventilation, n (%) | 36 (42) | 7 (32) | 8 (38) | 11 (52) | 10 (45) | 0.55 |
| Positive CAM score before first blood collection, n (%) | 10 (12) | 1 (5) | 4 (19) | 3 (14) | 2 (9) | 0.48 |
| Positive CAM score during the stay, n (%) | 19 (22) | 4 (18) | 7 (33) | 3 (14) | 5 (23) | 0.48 |
| Positive CAM score within the 96 hours of blood collection, n (%) | 7 (8) | 0 (0) | 4 (19) | 2 (10) | 1 (5) | 0.13 |

## **Discussion**

In this study, we successfully replicated the model proposed by Wittenbrink et al. for determining internal circadian time, achieving comparable high accuracy. We quantified the circadian desynchrony index by comparing the internal circadian time with the external blood sample collection time. Our results demonstrated that this index was significantly higher among critically ill ICU patients compared to healthy subjects. Notably, the majority of our ICU patients had a circadian desynchrony index greater than 9 hours. Additionally, the index was lower in the group where blood samples were drawn after 3pm. The distribution of clinical characteristics and outcomes across patient groups, categorized by the degree of circadian desynchrony, showed no statistical significance.

The significantly higher circadian desynchrony index among critically ill ICU patients compared to healthy subjects highlights the severe disruption of circadian rhythms in this population. This disruption can be attributed to numerous factors prevalent in the ICU environment, such as nightly disruption (e.g., care routines, feeding), constant exposure to artificial lighting, noise exposure and the stress associated with critical illness. The findings align with previous studies indicating that ICU settings can severely impact patients' circadian rhythms, potentially leading to adverse health outcomes. Excessive sound or light, surgical procedures, and mechanical ventilation have been reported to negatively affect sleep quality, thereby disrupting circadian rhythms in critically ill patients.[10-12] Impaired circadian rhythms have been reported in ICU patients and may result from ambient noise and the inconsistency of lighting conditions, including low lighting levels during the day and bright light at night.[13]

The symptoms of delirium are closely related to circadian rhythm and sleep disruption. In patients developing delirium, the sleep/wake cycle is reversed, with approximately 73% of delirious patients experiencing at least moderate alterations of sleep/wake cycle.[14,15] Several studies have reported that delirium is accompanied by a loss of melatonin circadian rhythms. For instance, Ángeles-Castellanos et al. investigated the association between circadian melatonin variation and delirium among hospitalized patients. They found that patients who did not develop delirium exhibited a daily melatonin rhythm. In contrast, those who developed delirium showed a loss of this rhythm, with mean melatonin levels decreasing as early as three days before the diagnosis of delirium. Additionally, Li et al. compared delirium and non-delirium patients during the first three days after admission to the surgical ICU.[16] They found that the non-delirium group exhibited a rhythmic pattern in melatonin and cortisol levels, whereas this rhythmicity was absent in the delirium group. In our study, although no statistically significant differences were identified in the rates of delirium across different levels of circadian desynchrony index during the blood collection period, the trends suggest that patients with higher circadian desynchrony indices may be more prone to delirium.

The study has several limitations. First, our collected data lacked a gold standard for internal circadian time. Although the model was developed and validated exclusively among healthy subjects, it remains to be determined whether it performs equally well in our critically ill patients. This could introduce bias into our findings. Second, the circadian desynchrony index may vary over time, and accurately assessing its impact on delirium status would require frequent sampling. Our collected data, however, included at most two time points, which may not capture the full variability of circadian desynchrony.

## **Conclusions**

In this work, we developed an approach to quantify circadian desynchrony index among ICU patients. The approach is simple and requires only a single blood sample taken any time during the day. While the rates of delirium varied across different levels of circadian desynchrony index, no statistically significant differences were identified. Nevertheless, the trends indicate that patients with higher circadian desynchrony indices may be more prone to delirium, highlighting the need for further research to explore these associations in more depth.